# CPT TESTS IN THE NEUTRAL KAON SYSTEM BY FNAL E773 *

Bernhard Schwingenheuer
University of Chicago
Chicago, Illinois 60637

**Abstract**

The FNAL experiment E773 has measured the phases $\Phi_{+-} = 43.35° \pm 0.70° \pm 0.79°$ and $\Phi_{00} - \Phi_{+-} = 0.67° \pm 0.85° \pm 1.1°$ of the CP violating parameters $\eta_{+-}$ and $\eta_{00}$ in the decay of neutral kaons into two charged or neutral pions. These preliminary results test CPT symmetry and show no evidence for a violation. In addition we present a preliminary measurement of $\Delta m = (0.5286 \pm 0.0029 \pm 0.0022) \cdot 10^{10}\ \hbar/sec$ and $\tau_S = (0.8929 \pm 0.0014 \pm 0.0014) \cdot 10^{-10}\ sec$. The first errors are statistical and the second errors are systematic for the above results.

For a long time it was thought that all interactions are invariant under the operations parity P, time reversal T and charge conjugation C. It was therefore a surprise when in 1957 three experiments [1, 2, 3] found that parity is violated in the weak interaction. Subsequently in 1964, another experiment found that the combination CP is also violated in weak decays of the neutral kaon at the $10^{-3}$ level [4].

For the combined operation CPT there is a strong theoretical prejudice stated in the CPT theorem [5]: every local field theory (e.g. the standard model) is CPT symmetric. To date, no violation of CPT has been reported.

Since the strangeness eigenstates $K^0$ and $\bar{K}^0$ can decay into the same final states the physical states are linear combinations:

$$K_{S,L} = \frac{1}{\sqrt{2(1+|\epsilon|^2)}}((1+\epsilon)K^0 \pm (1-\epsilon)\bar{K}^0). \quad (1)$$

CP violation in the neutral kaon decay is parameterized by the measurable ratio of violating to conserving amplitude:

$$\eta_{+-} = \frac{A(K_L \to \pi^+\pi^-)}{A(K_S \to \pi^+\pi^-)} = |\eta_{+-}|e^{i\Phi_{+-}}, \quad (2)$$

and a similar definition of $\eta_{00}$ for neutral pion decays. If CPT is not violated one can show that to within 0.1° [6]

$$\Phi_{+-} \approx \Phi_\epsilon \approx \Phi_{sw} \equiv \tan^{-1}\frac{2\Delta m}{\Delta\Gamma}, \quad (3)$$

*Talk given at the Conference on the Intersections of Particle and Nuclear Physics, St. Petersburg, Florida, June 1994.
E773 is a collaboration of the University of Chicago, Fermilab, Elmhurst College, University of Illinois and Rutgers University

$$\Delta\Phi = \Phi_{00} - \Phi_{+-} \approx 0. \tag{4}$$

with $\Delta m = m_L - m_S$ and $\Delta\Gamma = \Gamma_S - \Gamma_L$.

In E773 a beam of $K_L$ impinged on a regenerator consisting of plastic scintillator. After the regenerator the beam was a coherent superposition $|K_L\rangle + \rho|K_S\rangle$ with $\rho$ being the amplitude for forward regeneration (scattered kaons were treated as background). The decay rate into two pions as a function of proper time after the regenerator is then:

$$R(t) \propto |\eta e^{-it(m_L - i\Gamma_L/2)} + \rho e^{-it(m_S - i\Gamma_S/2)}|^2. \tag{5}$$

By fitting the decay distribution for neutral and charged pions the phases $\Phi_{+-}$ and $\Delta\Phi$ can be extracted from the interference term. In addition we can fit for the mass difference $\Delta m$ and the $K_S$ life time $\tau_S = 1/\Gamma_S$.

Experimentally, $\rho$ follows a power law [7, 8] as a function of kaon energy except for a correction due to decays of $K_S$ inside the regenerator ($\rho \sim p^\alpha g(p)$). If one further assumes that the elastic kaon-nucleon scattering amplitudes are analytic functions then the phase of $\rho$ can be related to the power law constant $\alpha$ using dispersion relations. Thus the regeneration amplitude can be parameterized by, say, its magnitude at 70 $GeV$ and $\alpha$.

The experiment E773 took data for three months in the 1991 fixed-target run at FNAL. The apparatus (Fig. 1) is essentially the same as for the 1987-88 run (experiment E731B [8]). Two $K_L$ beams are produced by a 800 $GeV/c$ proton beam striking a beryllium target. The detector components are located approximately 120 $m$ downstream of the target. While E731 used one regenerator in one of the beams, E773 ran with two regenerators, one at $z = 117.14$ $m$ (the upstream regenerator, UR) and one in the other beam at $z = 128.42$ $m$ (the downstream regenerator, DR). The UR is 1.2 and the DR is 0.4 interaction lengths long. Each of them toggled every minute between the two beams. Since the regenerators were constructed out of scintillator, i.e. active, inelastic interactions of kaons or neutrons in them could be vetoed.

The decay region extended from the regenerators to the end of the vacuum tank at $z = 159$ $m$. Inside the vacuum tank and throughout the spectrometer were photon vetoes at 11 different $z$ positions to detect charged and neutral particles leaving the fiducial region of the detector. Following the vacuum tank was a charged spectrometer consisting of four drift chambers and a magnet, which was used to find charged tracks and to measure their momenta. The momentum resolution was $\sigma_p/p = 0.45\,(1 \oplus p/40)\%$ with $p$ in units of $GeV/c$. After the spectrometer there were two scintillator planes with horizontal or vertical segmentation. The trigger for the two-body charged decays $K \to \pi^+\pi^-$ required up-down and left-right hits in the hodoscopes and drift chambers. For part of the run there was another trigger hodoscope at $z = 141$ $m$ which was removed, after a trigger upgrade, to extend the decay region for charged decays.

Following the hodoscope was a circular array of 804 lead glass blocks (each block 5.8 $cm$ by 5.8 $cm$ and 18.7 $X_0$ deep) with holes in the middle of the array for the two beams. This was used as an electromagnetic calorimeter for the neutral decays $K \to \pi^0\pi^0$. The trigger required at least 25 $GeV$ in the calorimeter and 4 or 6 clusters found by a hardware cluster finder.

All electromagnetic shower energy was contained in the lead glass and the lead wall (21.4 $X_0$) downstream of it. A scintillator plane further downstream was therefore used to veto hadronic showers. Downstream of this plane were 3 $m$ of iron and another scintillator plane. Muons from $K \rightarrow \pi\mu\nu$ decays would penetrate this material and could thus be detected and vetoed in the charged mode trigger.

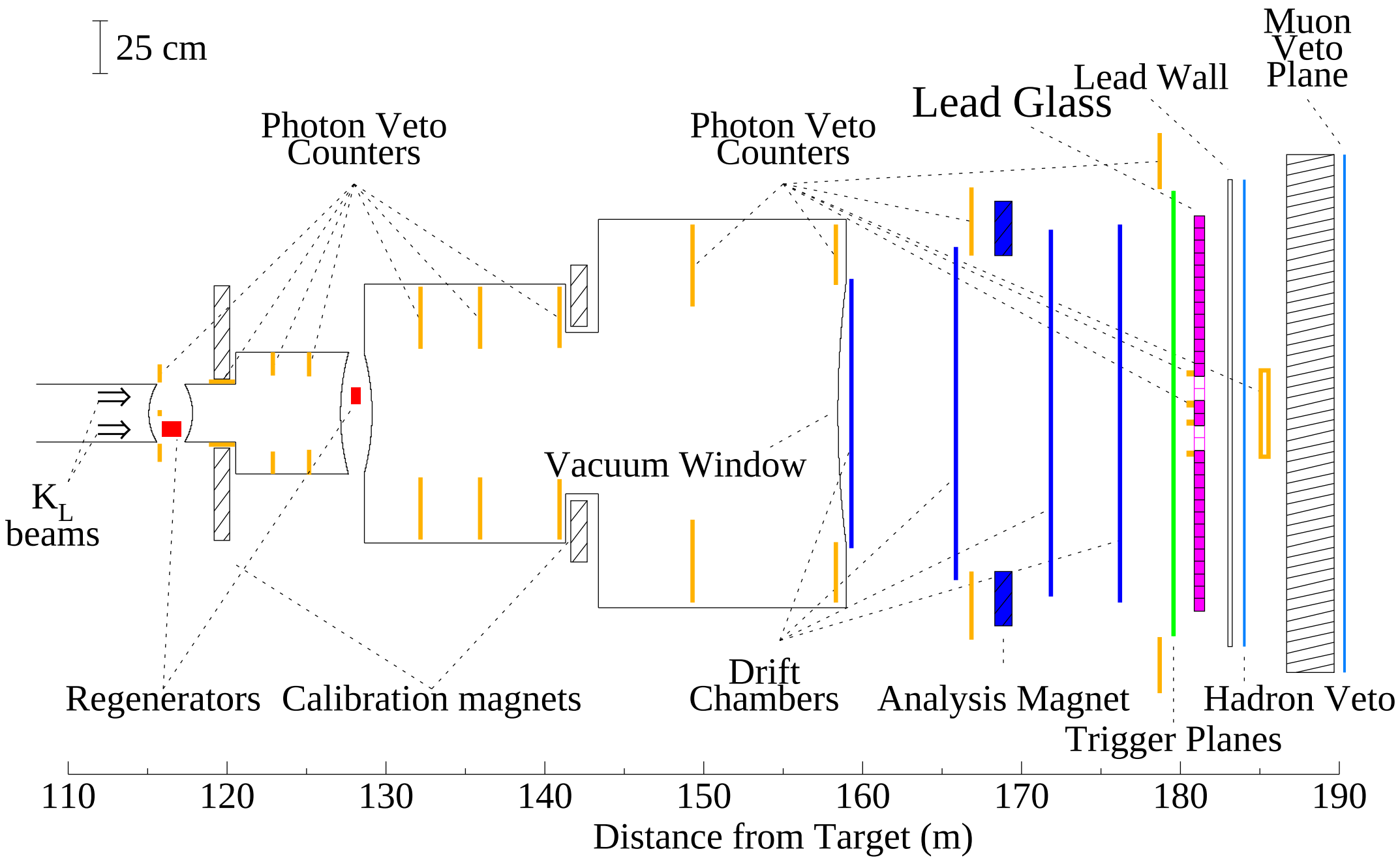

Figure 1: Schematic detector picture of experiment FNAL E773.

Drift chamber calibration constants (transverse position and rotation of chambers and time-to-distance relations) were generated about once a day by using tracks from special runs with the magnet off and "two-track" decays. For the lead glass calibration a large sample of electrons from $K \rightarrow \pi e\nu$ decays was used. Because of increasing radiation damage throughout the run the opacity of the glass changed. Blocks that were hit most were therefore calibrated about once a day. The gain of the photo tube/ADC and the light absorption coefficient for every block was determined.

The lead glass was calibrated using electrons. Because of absorption the light output of a photon shower at a given energy was different from that of an electron shower. To correct for the size of the electron-photon cluster differences we relied on a simulation which was fine tuned by comparing the Monte Carlo and data $z$ distributions of $K \rightarrow \pi^0\pi^0$ and $K \rightarrow 3\pi^0$ around the regenerators. The $z$ position of this sharp edge was sensitive to small changes in the energy scale and a small energy dependent correction (about 0.3%) was applied.

To reconstruct $K \rightarrow \pi^+\pi^-$ events, two tracks were found in the $x$ and $y$ view and then matched to clusters of energy in the lead glass. The ratio of cluster energy to spectrometer momentum $E/p$ had to be less than 0.8 to reject electrons from semileptonic decays $K \rightarrow \pi e\nu$. Track and vertex quality cuts as well as aperture cuts were applied. A minimum momentum requirement ensures that a muon would have penetrated the muon filter and fired the muon veto. Events with a $p\pi$ mass within 6 $MeV/c^2$ of the $\Lambda$ mass were rejected to reduce $\Lambda \rightarrow p\pi$ background. $K \rightarrow \pi^+\pi^-\pi^0$ events were rejected by requiring the $\pi^+\pi^-$ mass to be close to the kaon mass. Finally, kaons that scattered in the regenerator with a transverse momentum $p_T^2 > 250\ (MeV/c)^2$ were discarded. After all cuts the background was quite small (0.3% and 0.8% for the UR and DR beam, respectively).

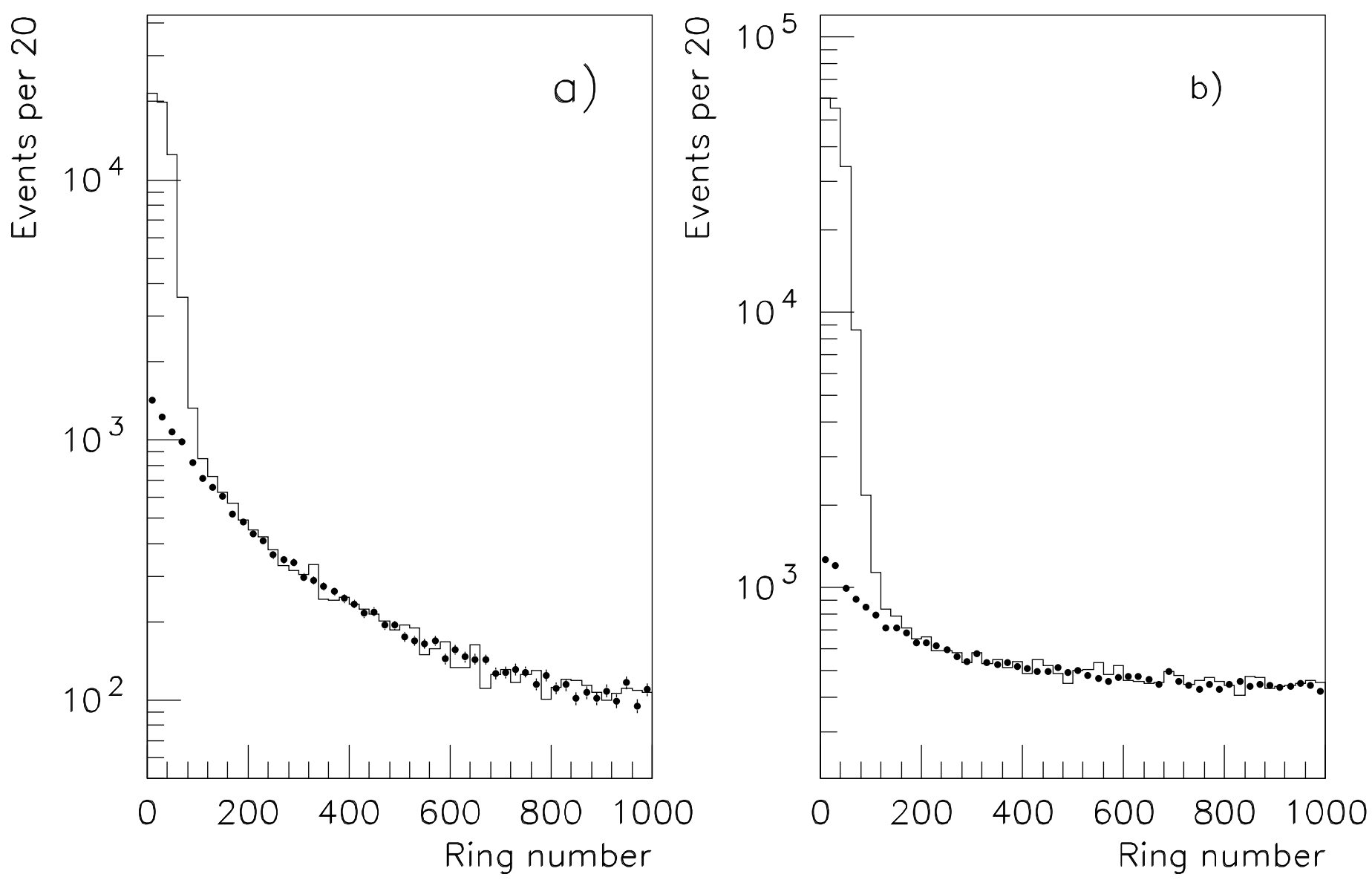

Figure 2: Ring number distributions of $K \rightarrow \pi^0\pi^0$ for the downstream (part a) and upstream regenerator (part b). Shown are the data (histogram) and a Monte Carlo simulation of the background from scattered kaons(circles). The cut was at 120.

For $K \rightarrow \pi^0\pi^0$ decays the four clusters in the lead glass were combined into two pairs. Assuming two $\pi^0$ decays the $z$ decay vertices of the pions were calculated. With four photons there are three combinations. If the two $z$ values agreed with a $\chi^2 < 4$ the event was kept. Further requirements were: four cluster mass between 474 $MeV/c^2$ and 522 $MeV/c^2$, off-line cut on photon veto energies, and cut on the shape of the electromagnetic cluster (to ensure that each cluster came from a single photon). All three cuts reduced the $K \rightarrow \pi^0\pi^0\pi^0$ background. Events with out-of-time energy due to accidental activity were rejected. The four photon center of energy was the position where the kaon would have hit the calorimeter if it had not decayed. The maximum distance between this position and the center of the beam in the $x$ or $y$ view (called "ring number") had to be small. For a kaon that scattered in a regenerator it was likely that its position at the glass was outside the beam profile and would therefore be rejected.

The background to the neutral mode came from three different sources. For $K_L \rightarrow \pi^0\pi^0\pi^0$ with two photons escaping detection or fusing, a Monte Carlo simulation was used. The normalization was taken from the mass side bands since all events in the data that reconstruct outside the kaon mass and do not come from beam interactions, i.e. do not reconstruct at the $z$ positions of vacuum windows, were $K \rightarrow 3\pi^0$ decays. The background fractions were 0.6%(1.4%) for the UR(DR) beam.

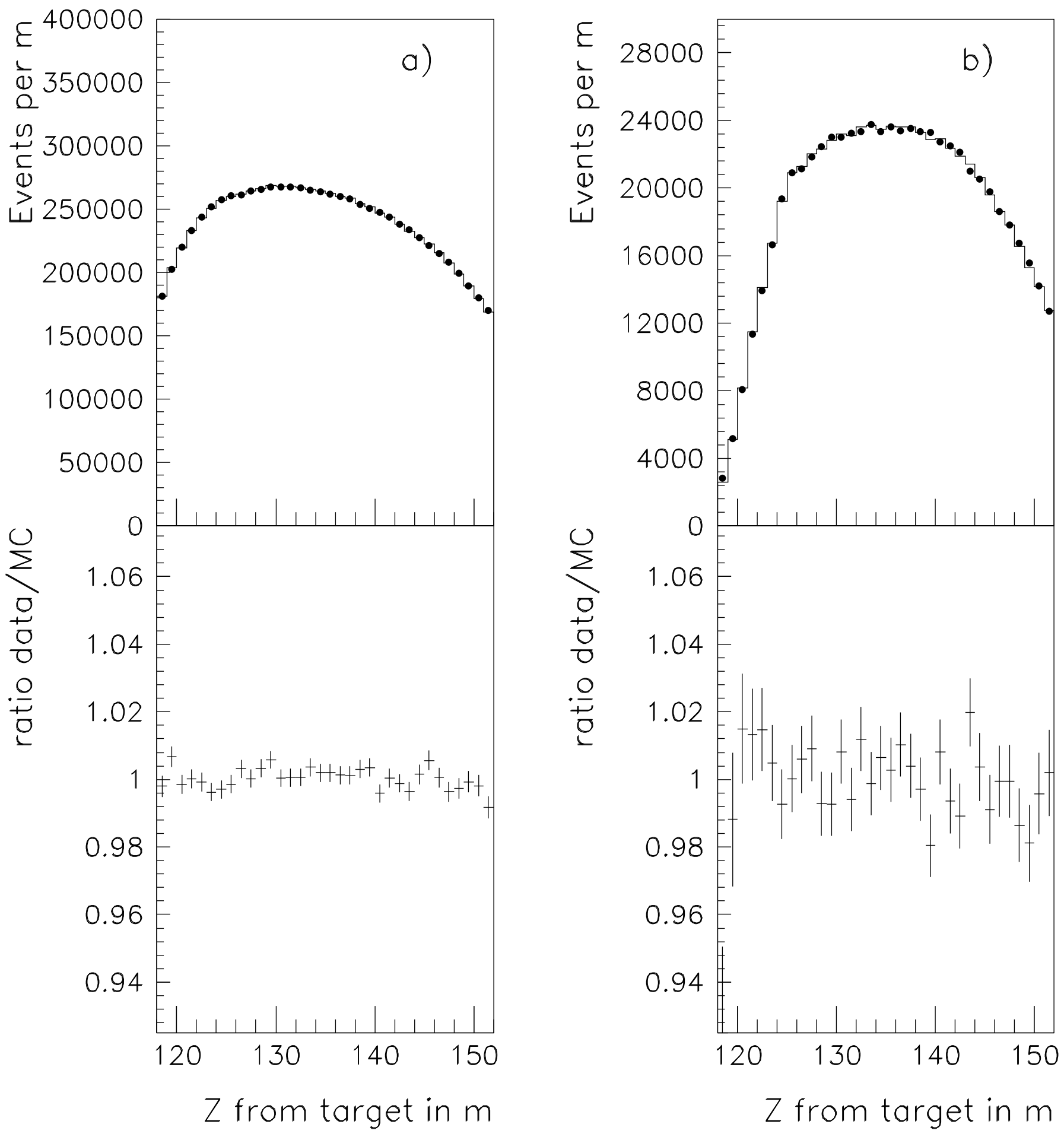

Figure 3: Decay distribution of the upstream regenerator beam (a) $K_L \to \pi e \nu$ and (b) $K_L \to 3\pi^0$ decays. The upper plots show data (histogram) and Monte Carlo (circles) while the lower plots show the ratio of data over Monte Carlo.

The background from beam interactions in the vacuum windows or air around the regenerators and the regenerators itself was subtracted by linear interpolation in the four photon mass distribution. The size of this background was about 0.1%.

The third class of background was due to scattered kaons. To estimate the amount of background that was left after the ring number cut a Monte Carlo simulation of the scattering was used. The input to the simulation was taken from the $K \to \pi^+\pi^-$ decays, since the kaon $p_T^2$ could be calculated for these events. The kaon could either scatter inelastically which results in a rather flat $p_T^2$ distribution ($\exp(-5p_T^2)$ with $p_T^2$ in units of $GeV^2/c^2$) or diffractively. The diffractive $K_L \to K_L$ scattering was parameterized by $\exp(-58p_T^2)$ while the diffractive $K_L \to K_S$ scattering was parameterized by $\exp(-90p_T^2)$. The exponents were the same for both regenerators. Together with two normalizations, one for each regenerator, the diffractive background was thus parameterized by four constants in the Monte Carlo which predicted the $z$, energy and $p_T^2$ distributions for all decay modes quite successfully. Fig. 2 shows the ring number distribution for $\pi^0\pi^0$ decays and the Monte Carlo simulation of the background. The background fractions were 3.0% for the upstream (thicker) and 9.7% for the downstream (thinner) regenerator.

Understanding the acceptance of the detector was crucial for this experiment since the reported measurements are obtained by fitting the acceptance corrected decay distributions to the expected rate (5). The acceptance was checked by comparing high statistics data and Monte Carlo $z$ distributions of $K_L$ decays into $3\pi^0$ and $\pi e\nu$ for the $\pi^0\pi^0$ and $\pi^+\pi^-$ decay mode, respectively. These decay modes were not sensitive to the simulation of the regeneration. Fig. 3 shows that the acceptance is understood to within 1%.

Fig. 4 shows as an example the corrected distribution of $K \to \pi^+\pi^-$ for one kaon energy bin. Three different kinds of fits to the background subtracted and acceptance corrected decay distributions were performed. In all fits the two quantities parameterizing the regeneration amplitude $\rho$ ($|\rho(70\,GeV)|$ and $\alpha$), normalizations for the kaon flux and two shape parameters for its energy dependence were floated. In the first fit $\Phi_{+-}$ was floated in addition. In the fit for $\Delta m$ and $\tau_S$, $\Phi_{+-}$ was fixed to $\Phi_{sw}$ (assuming CPT invariance, eqn. 3). For $\Delta\Phi$, both phases, $|\epsilon'/\epsilon|$ ($3\epsilon' = \eta_{+-} - \eta_{00}$), a normalization for the relative number of $\pi^0\pi^0$ to $\pi^+\pi^-$ events and two more energy dependend parameters for the $K \to \pi^0\pi^0$ flux [1] were floated.

The fit result for $\Phi_{+-}$ and its dependence on $\Delta m$ and $\tau_S$ (for one Particle Data Group (PDG) standard deviation [9]) is:

$$\begin{aligned}\Phi_{+-} = 43.35^\circ \; &\pm\; 0.70^\circ(stat) \;\pm\; 0.79^\circ(syst) \\ &-\; 0.62^\circ\,\frac{\tau_S - 0.8922\cdot 10^{-10}sec}{0.0020\cdot 10^{-10}sec} + 0.38^\circ\,\frac{\Delta m - 0.5286\cdot 10^{10}\hbar/sec}{0.0024\cdot 10^{10}\hbar/sec}.\end{aligned} \quad (6)$$

Note that the central value for $\tau_S$ is taken from PDG but that the $\Delta m$ value is our best fit which is motivated by the argument described below.

The systematic errors for the fits are listed in table 1. For the $\Delta\Phi$ fit the systematic error due to the minimal cluster energy cut as mentioned below is part of the systematic due to the nonlinear lead glass response.

In the $\tau_S$ and $\Delta m$ fit we get:

$$\Delta m = (0.5286 \pm 0.0029(stat) \pm 0.0022(syst))\cdot 10^{10}\hbar/sec, \quad (7)$$

$$\tau_S = (0.8929 \pm 0.0014(stat) \pm 0.0014(syst))\cdot 10^{-10}sec. \quad (8)$$

The result for $\Delta m$ is very interesting since it deviates from the PDG value $\Delta m_{PDG} = (0.5351 \pm 0.0024)\cdot 10^{10}\hbar/sec$ [9], but confirms the most recent measurement from E731 [13]. This is relevant for the CPT test $\Phi_{+-} \approx \Phi_{sw}$ since the value of $\Phi_{sw}$ depends on $\Delta m$ and because there is a strong correlation between $\Delta m$ and $\Phi_{+-}$ in the phase fit. This is true for all published measurements. The current PDG values are $\Phi_{+-,PDG} = 46.6^\circ \pm 1.2^\circ$ and $\Phi_{sw,PDG} = 43.73^\circ \pm 0.15^\circ$ which hints at CPT violation. If however previous experimental results are corrected for their $\Delta m$ dependence to our lower value of $\Delta m$ the discrepancy vanishes. This is shown in Fig. 5. The open circles are the results used to determine $\Phi_{+-,PDG}$ while the solid points are the corrected measurements and more recent results (solid squares). The hatched band shows $\Phi_{sw}$ using our central value for $\Delta m$. Note that after the correction there is perfect agreement between $\Phi_{sw}$ and $\Phi_{+-}$ and among all experiments.

[1] Because of acceptance effects not simulated in the Monte Carlo there was a small mismatch in the energy spectrum for $\pi^0\pi^0$ and $\pi^+\pi^-$ decays.

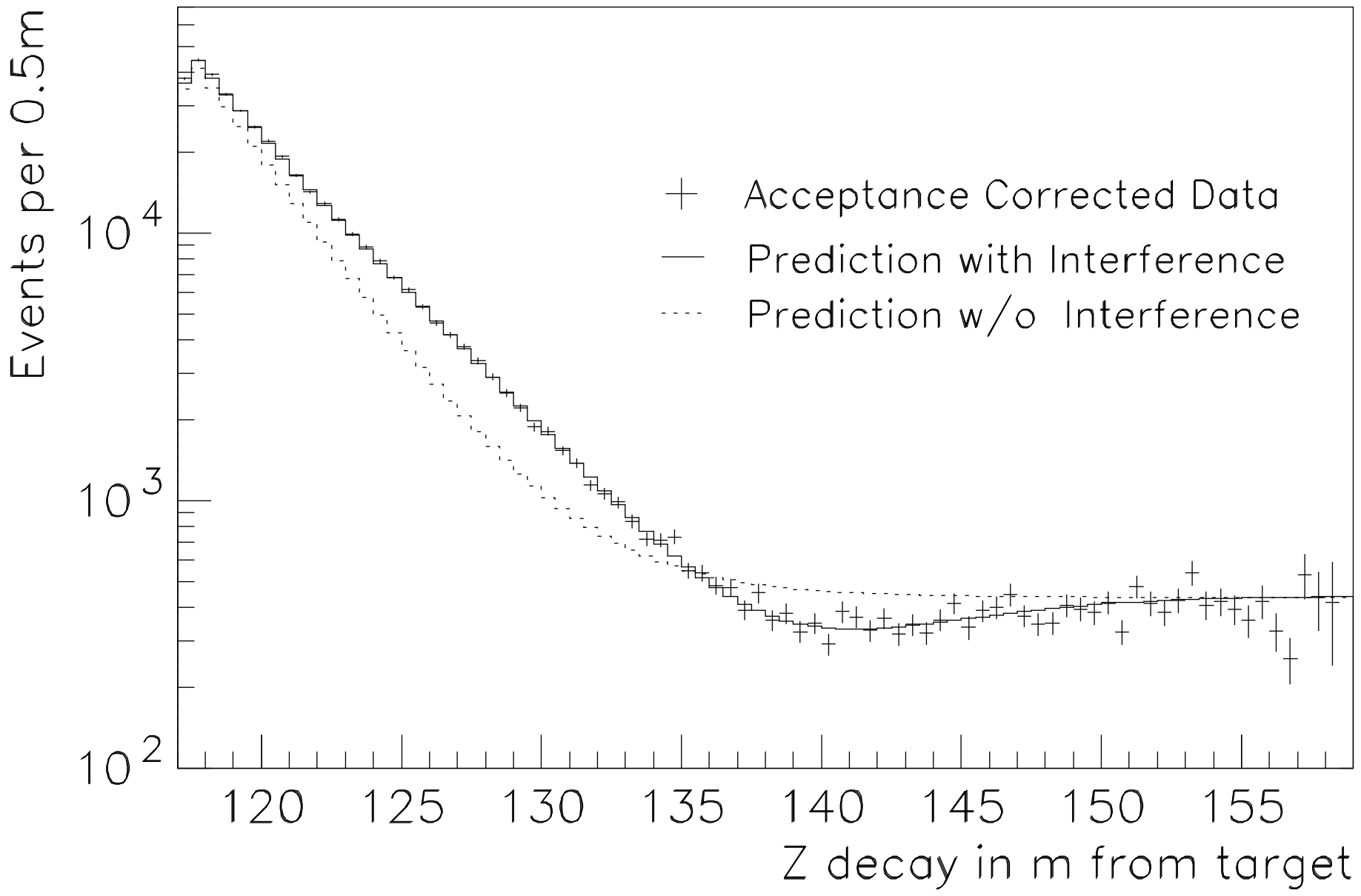

Figure 4: Background subtracted and acceptance corrected decay distribution of $K \to \pi^+\pi^-$ for the upstream regenerator beam in the energy bin $40 - 50$ $GeV$ (crosses) and the predictions from the fit with and without the interference term.

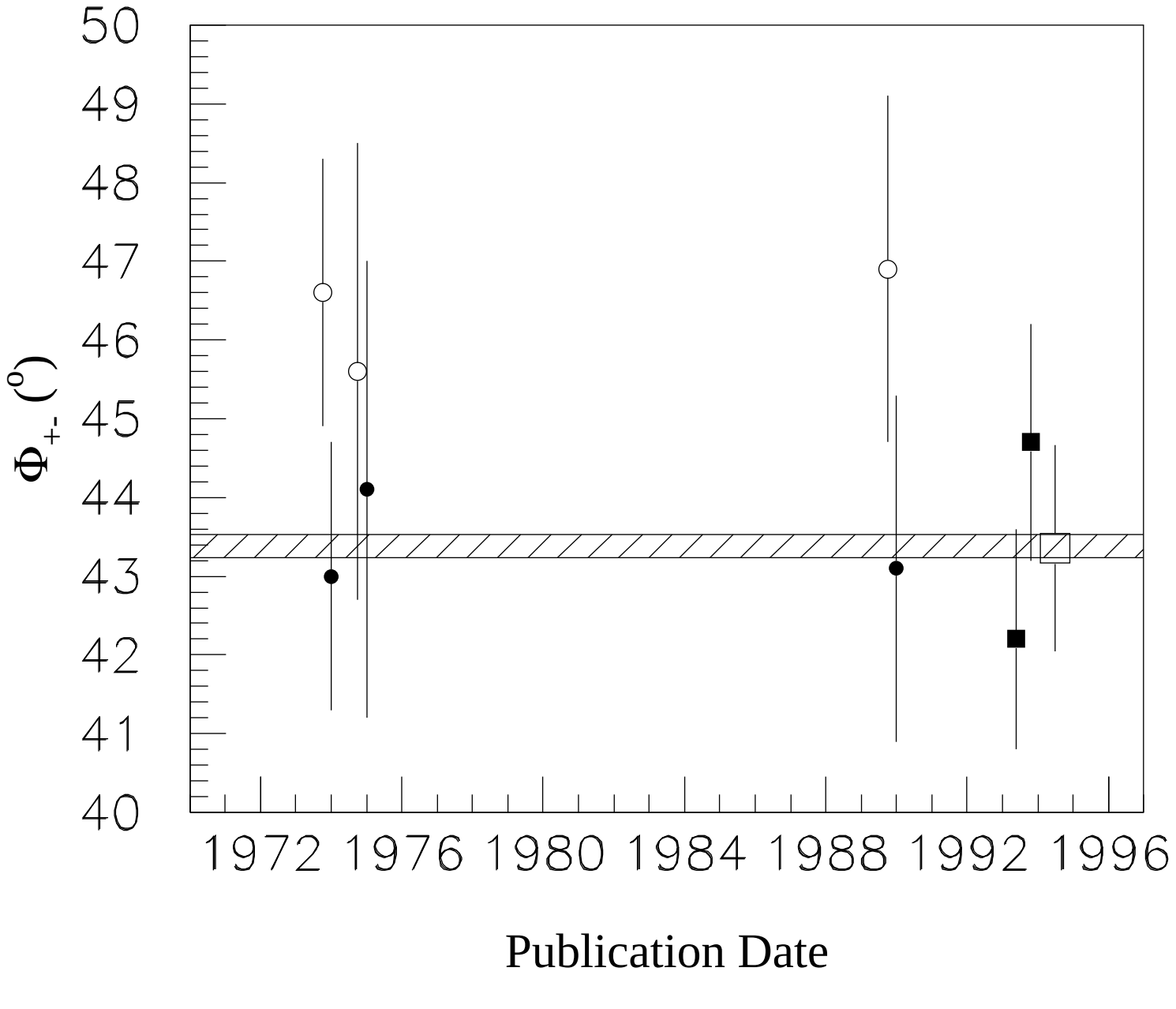

Figure 5: Previous measurements of $\Phi_{+-}$ and this result (open square). The open circles are the results used to determine $\Phi_{+-,PDG}$, the solid points are the same measurements corrected using our $\Delta m$ and the solid squares are more recent results. The band is the $\pm 1$ $\sigma$ band of $\Phi_{sw}$. The corrected values all agree with each other and with $\Phi_{sw}$. Thus the test $\Phi_{+-} \sim \Phi_{sw}$ shows no hint of CPT violation anymore. The $\tau_S$ and $\Delta m$ dependences are included in the errors. The references to the results are in time order [10, 11, 12, 13, 14].

Table 1: Systematic error estimates for the fits. The individual sources were added in quadrature to give the total error.

| | $\Phi_{+-}$ fit | $\Delta m/\tau_S$ fit | | $\Delta\Phi$ fit |
|---|---|---|---|---|
| | in degree | in $10^{10}\hbar/sec$ | in $10^{-10}sec$ | in degree |
| regenerator positions, lengths | 0.1 | 0.0008 | 0.0004 | - |
| charged mode acceptance | 0.57 | 0.0017 | 0.0010 | 0.2 |
| neutral mode acceptance | - | - | - | 0.2 |
| background subtraction | 0.2 | 0.0005 | 0.0002 | 0.6 |
| deviations from the power law for the reg. amplitude | 0.5 | 0.0011 | 0.0008 | - |
| lead glass energy scale | - | - | - | 0.6 |
| lead glass energy resolution | - | - | - | 0.5 |
| lead glass nonlinear response | - | - | - | 0.5 |
| total | 0.79 | 0.0022 | 0.0014 | 1.1 |

The preliminary $\Delta m$ result from CPLEAR, presented at this conference [15], is $\Delta m = (0.5318 \pm 0.0044(stat) \pm 0.0010(syst) \pm 0.0030(MC)) \cdot 10^{10}\ \hbar/sec$, consistent with our and E731 measurements.

For the $\Delta\Phi$ fit we used 70% of the data for the preliminary analysis. The fit result was sensitive to the cut on the minimal cluster energy in the lead glass. Changing the cut from 2.2 $GeV$ to 4.0 $GeV$ changed the result by 0.9° which is not yet understood. For the preliminary result we have decided to take the average of the two fits as the central value and to assign 0.5° as a systematic error due to this effect. The preliminary result is:

$$\Delta\Phi = 0.67^\circ \pm 0.85^\circ(stat) \pm 1.1^\circ(syst). \tag{9}$$

We find that $\Delta\Phi$ is consistent with zero (eqn. 4) and our values of $\Phi_{sw}$ and $\Phi_{+-}$ agree (eqn. 3). Therefore both tests find that CPT symmetry is valid to the accuracy of our experiment.

In addition to the CPT tests, E773 has the best measurements of the CP violation parameter in the radiative decay $K \to \pi^+\pi^-\gamma$. Our preliminary results are:

$$|\eta_{+-\gamma}| = (2.414 \pm 0.065(stat) \pm 0.062(syst)) \cdot 10^{-3}, \tag{10}$$

$$\Phi_{+-\gamma} = 45.5^\circ \pm 3.6^\circ(stat) \pm 2.4^\circ(syst). \tag{11}$$

Thus $\eta_{+-\gamma}$ for the radiative decay (inner Bremsstrahlung) is consistent with being the same as $\eta_{+-}$ [9]. These measurements establish an improvement in precision by more than a factor of 3 relative to the previous results [16].